\documentclass[aps,amsmath,amssymb,twocolumn,floats]{revtex4-1}

\setcitestyle{super}
\usepackage{graphicx}
\begin{document}

\title{Marginal evidence for cosmic acceleration from Type Ia supernovae}

\author{J. T. Nielsen$^1$}
\author{A. Guffanti$^2$}
\author{S. Sarkar$^{1,3}$}

\affiliation{$^1$Niels Bohr International Academy, Blegdamsvej 17, Copenhagen 2100, Denmark}
\affiliation{$^2$ Universitˆ degli Studi di Torino, via P. Giuria 1, I-10125 Torino, Italy}
\affiliation{$^3$Rudolf Peierls Centre for Theoretical Physics, 1 Keble Road, Oxford OX1 3NP, UK}

\begin{abstract}
  The `standard' model of cosmology is founded on the basis that the
  expansion rate of the universe is accelerating at present --- as was
  inferred originally from the Hubble diagram of Type Ia
  supernovae. There exists now a much bigger database of supernovae so
  we can perform rigorous statistical tests to check whether these
  `standardisable candles' indeed indicate cosmic acceleration. Taking
  account of the empirical procedure by which corrections are made to
  their absolute magnitudes to allow for the varying shape of the
  light curve and extinction by dust, we find, rather surprisingly,
  that the data are still quite consistent with a constant rate of
  expansion.
\end{abstract}

\maketitle

\section{Introduction}

In the late 1990's, studies of Type Ia supernovae (SN Ia) showed that
the expansion rate of the universe appears to be accelerating as if
dominated by a cosmological
constant~\cite{Perlmutter:1998np,Riess:1998cb,Goobar:2011iv}. Since
then supernova cosmology has developed rapidly as an important probe
of `dark energy'. Empirical corrections are made to reduce the scatter
in the observed magnitudes by exploiting the observed
(anti)correlation between the peak luminosity and the light curve
width~\cite{Phillips:1993ng,Tripp:1997wt}. Other such correlations
have since been found e.g. with the host galaxy
mass~\cite{Kelly:2009iy} and
metallicity~\cite{Hayden:2012aa}. Cosmological parameters are then
fitted, along with the parameters determining the light curves, by
simple $\chi^2$
minimisation~\cite{Perlmutter:1998np,Astier:2005qq,Conley:2006qb,Kowalski:2008ez,Betoule:2014frx}. This
method has a number of pitfalls as has been emphasised
earlier~\cite{Vishwakarma:2010nc,March:2011xa}.

With ever increasing precision and size of SN Ia datasets, it is
important to also improve the statistical analysis of the data. To
accomodate model comparison, previous
work~\cite{Kim:2011hg,Lago:2011pk,Wei:2015xca} has introduced
likelihood maximisation. In this work we present an improved maximum
likelihood analysis, finding rather different results.

\section{\label{sec:snsalt}Supernova cosmology}

There are several approaches to making SN Ia `standardiseable
candles'. The different philosophies lead to mildly different results
but the overall picture seems consistent~\cite{Hicken:2009dk}. In this
paper we adopt the transparent approach of `Spectral Adaptive
Lightcurve Template 2' (SALT2)~\cite{Guy:2005me,Guy:2007dv} wherein
the SN Ia are standardised by fitting their light curve to an
empirical template, and the parameters of this fit are used in the
cosmological analysis. Every SN Ia is assigned three parameters, one
being $m^*_B$, the apparent magnitude at maximum (in the rest frame
`$B$-band'), while the other two describe the light curve shape and
colour corrections: $x_1$ and $c$. The distance modulus is then taken
to be:
\begin{equation}
\label{MUSN}
\mu_\textrm{SN} = m^*_B - M + \alpha x_1 - \beta c,
\end{equation}
where $M$ is the absolute magnitude, and $\alpha$ and $\beta$ are
assumed to be constants for \emph{all} SN Ia. These global constants
are fitted along with the cosmological parameters. The physical
mechanism(s) which give rise to the correlations that underlie these
corrections remain uncertain~\cite{Hoflich:1996sx,Kasen:2006is}.  The
SN Ia distance modulus is then compared to the expectation in the
standard $\Lambda$CDM cosmological model:
\begin{eqnarray}
&& \mu \equiv 25 + 5 \log_{10}(d_\textrm{L}/\textrm{Mpc}), 
 \quad \textrm{where:} \nonumber \\
&& d_\textrm{L} = (1 + z) \frac{d_\textrm{H}}{\sqrt{\Omega_k}} 
 \textrm{sinh}\left(\sqrt{\Omega_k} \int_0^z \frac{H_0 \textrm{d}z'}{H(z')}\right) ,
 \nonumber \\
&& d_\textrm{H} = c/H_0, \quad H_0 \equiv 
 100h~\textrm{km}\,\textrm{s}^{-1}\textrm{Mpc}^{-1}, \nonumber \\
&& H = H_0 \sqrt{\Omega_\textrm{m} (1 + z)^3 + \Omega_k (1 + z)^2 
 + \Omega_\Lambda},
\label{DLEQ}
\end{eqnarray}
where $d_\textrm{L}, d_\textrm{H}, H$ are the luminosity distance,
Hubble distance and Hubble parameter respectively, and
$\Omega_\textrm{m}, \Omega_\Lambda, \Omega_k$ are the matter,
cosmological constant and curvature density in units of the critical
density~\cite{Goobar:2011iv}.  There is a degeneracy between $H_0$ and
$M_0$ so we fix the value of the Hubble parameter today to $h = 0.7$
which is consistent with independent measurements.

\section{\label{sec:mle}Maximum Likelihood Estimators}
 
To find the maximum likelihood estimator (MLE) from the data, we must
define the appropriate likelihood:
\begin{equation}
\mathcal{L} = \textrm{probability density}(\textrm{data}|\textrm{model})
\nonumber,
\label{LIKELIHOOD}
\end{equation}
i.e. we have to first specify our model of the data. For a given SN
Ia, the true data $(m^*_B, x_1, c)$ are drawn from some global
distribution. These values are contaminated by various sources of
noise, yielding the observed values
$(\hat{m}^*_B,\hat{x}_1, \hat{c})$. Assuming the SALT2 model is
correct, only the true values obey equation \eqref{MUSN}. However when
the experimental uncertainty is of the same order as the intrinsic
variance as in the present case, the observed value is \emph{not} a
good estimate of the true value. Parameterising the cosmological model
by $\theta$, the likelihood function can be written as:
\begin{eqnarray} \label{LIKEINT}
{\mathcal{L}} 
 && = p [(\hat{m}^*_B, \hat{x}_1, \hat{c}) | \theta] \\
 && = \int p [(\hat{m}^*_B, \hat{x}_1, \hat{c}) | (M, x_1, c), \theta] \ 
  p [(M, x_1, c) | \theta] \textrm{d}M \textrm{d}x_1 \textrm{d}c, \nonumber
\end{eqnarray}
which shows explicitly where the experimental uncertainties enter
(first factor) and where the variances of the intrinsic distributions
enter (second factor).

\begin{figure}[!t]
    \includegraphics[width=\columnwidth]{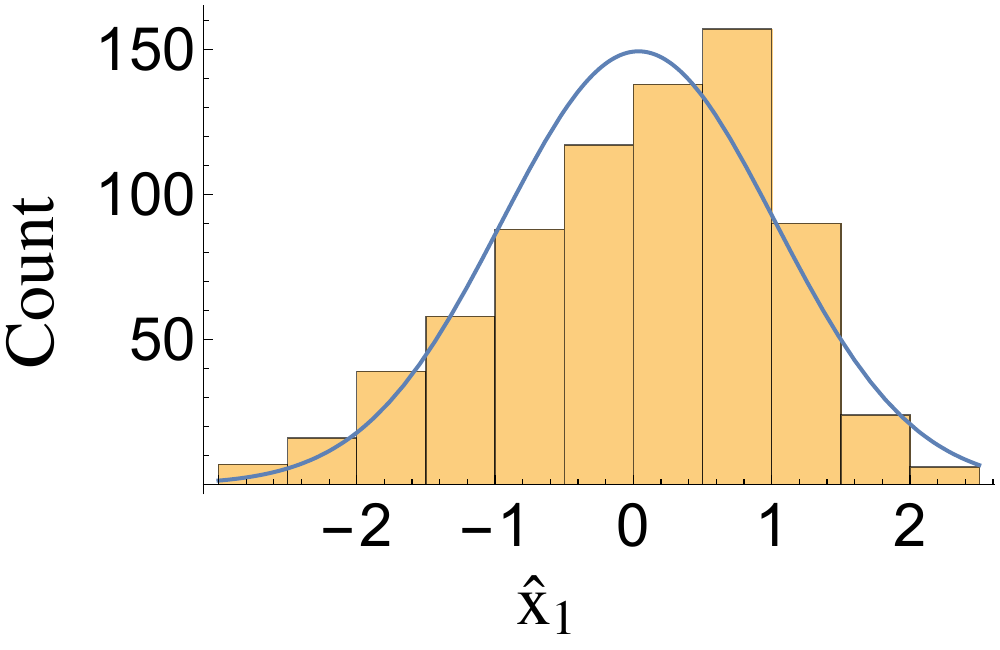}
    \includegraphics[width=\columnwidth]{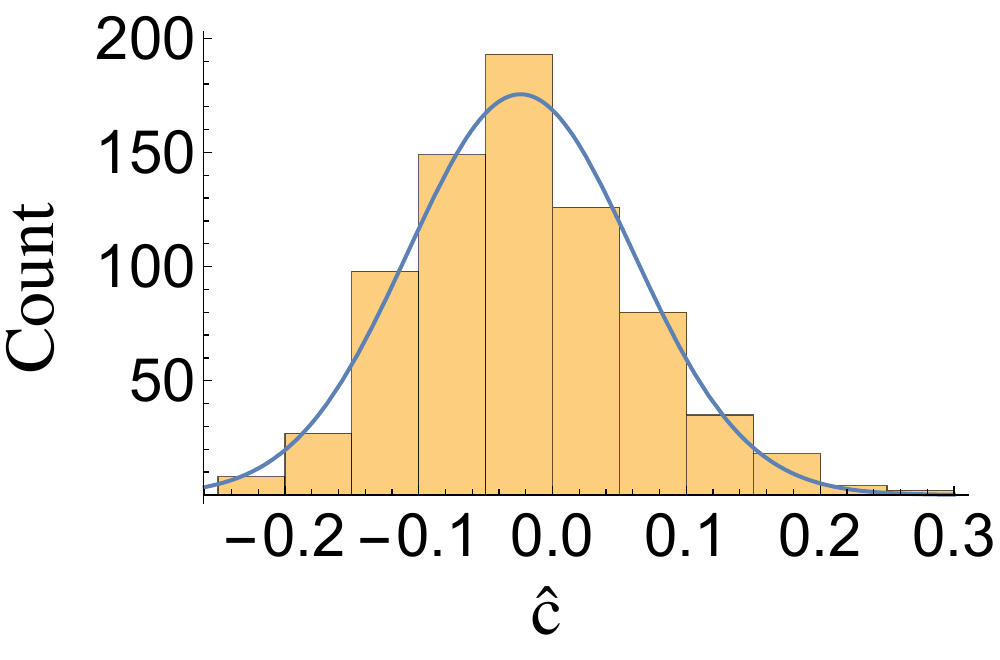}
    \caption{Distribution of the stretch and colour correction
      parameters in the JLA sample~\cite{Betoule:2014frx}, with
      gaussians superimposed.}
\label{fig:x1_c}
\end{figure}

Having a theoretically well-motivated distribution for the light curve
parameters would be helpful, however this is not available. For
simplicity we adopt global, independent gaussian distributions for all
parameters, $M, x_1$ and $c$ (see Fig.~\ref{fig:x1_c}), i.e. model
their probability density as:
\begin{eqnarray}
&& p [(M, x_1, c)|\theta] = p (M|\theta) p(x_1|\theta) p(c| \theta), 
 \quad \textrm{where:}\nonumber \\
&& p (M|\theta) = (2\pi\sigma_{M_0}^2)^{-1/2} 
 \exp\left\{-\left[\left({M - M_0}\right)/\sigma_{M_0}\right]^2/2\right\},
 \nonumber \\
 && p (x_1|\theta) = (2\pi\sigma_{x_{1,0} }^2)^{-1/2} 
 \exp\left\{-\left[\left({x_1 - x_{1,0} }\right)/\sigma_{x_{1,0} }\right]^2/2\right\}, 
 \nonumber \\
 && p (c|\theta) = (2\pi\sigma_{c_0}^2)^{-1/2} 
 \exp\left\{-\left[\left({c - c_0}\right)/\sigma_{c_0}\right]^2/2\right\}.
\label{MODELGAUSS}
\end{eqnarray}
All 6 free parameters $\{M_0,\sigma_{M_0}, x_{1,0}, \sigma_{x_{1,0} }, c_0, 
\sigma_{c_0}\}$ are fitted along with the cosmological
parameters and we include them in $\theta$. 
Introducing the vectors $Y= \{M_1, x_{11}, c_1, \dots
M_N, x_{1N}, c_N \}$, the zero-points $Y_0$, and the matrix $\Sigma_l =
\textrm{diag}(\sigma_{M_0}^2,\sigma_{x_{1,0} }^2,\sigma_{c_0}^2,\dots)$, the
probability density of the true parameters writes:
\begin{equation}
  p (Y|\theta) = {|2\pi\Sigma_l|}^{-1/2} 
  \exp\left[-(Y - Y_0) \Sigma_l^{-1} (Y - Y_0)^\textrm{T}/2\right],
\end{equation}
where $|\ldots|$ denotes the determinant of a matrix. What remains is
to specify the model of uncertainties on the data. Introducing another
set of vectors $X = \{m^*_{B1}, x_{11}, c_1, \dots\}$, the observed
$\hat X$, and the estimated experimental covariance matrix
$\Sigma_\textrm{d}$ (including both statistical and systematic
errors), the probability density of the data given some set of true
parameters is:
\begin{equation}
p (\hat X|X, \theta) = {|2\pi \Sigma_\textrm{d}|}^{-1/2}
 \exp \left[-(\hat{X} - X) \Sigma_\textrm{d}^{-1} (\hat{X} - X)^\textrm{T}/2 \right].
\end{equation}
To combine the exponentials we introduce the vector
$\hat Z = \{\hat m^*_{B1} - \mu_{1}, \hat x_{11}, \hat c_1, \dots \}$
and the block diagonal matrix
\begin{eqnarray}
A = \begin{pmatrix}
1 && 0 & 0 & \\
-\alpha && 1 & 0 & 0 \\
\beta && 0 & 1&  \\
 && 0 & \phantom{-1} & \ddots
\end{pmatrix}. \\
\nonumber 
\end{eqnarray}
With these, we have $\hat{X} - X = (\hat{Z}A^{-1} - Y) A$ and so
$p(\hat{X} | X, \theta) = p(\hat{Z} | Y, \theta)$. The likelihood is
then
\begin{eqnarray}
\mathcal L  
&&= \int p(\hat{Z} | Y, \theta)\ p(Y | \theta) \textrm{d}Y  \\
&&= {|2\pi \Sigma_\textrm{d}|}^{-1/2} {|2\pi\Sigma_l|}^{-1/2}
	 \int\textrm{d}Y  \nonumber \\	
&&\times\exp\left(-(Y-Y_0) \Sigma_l^{-1} (Y-Y_0)^\textrm{T}/2 \right) \nonumber \\
&&\times\exp\left(-(Y-\hat{Z} A^{-1}) A \Sigma_d^{-1} 
  A^\textrm{T}(Y - \hat{Z} A^{-1})^\textrm{T}/2 \right), \nonumber
\end{eqnarray}
which can be integrated analytically to obtain:
\begin{eqnarray}
{\mathcal L} 
&& = {|2\pi(\Sigma_\textrm{d} + A^\textrm{T}\Sigma_l A)|}^{-1/2} \\
&& \times \exp\left[-(\hat{Z} - Y_0 A)(\Sigma_\textrm{d} + A^\textrm{T}
 \Sigma_l A)^{-1}(\hat{Z} - Y_0 A)^\textrm{T}/2 \right].\nonumber 
\label{LIKEJUMP}
\end{eqnarray}
This is the likelihood (equation \eqref{LIKELIHOOD}) for the simple
model of equation \eqref{MODELGAUSS}, and the quantity which we
maximise in order to derive confidence limits. The 10 parameters we
fit are
$\{\Omega_\textrm{m}, \Omega_\Lambda, \alpha, x_{1,0}, \sigma_{x_{1,0}
}, \beta, c_0, \sigma_{c_0}, M_0, \sigma_{M_0} \}$.
We stress that it is necessary to consider all of these together and
$\Omega_\textrm{m}$ and $\Omega_\Lambda$ have no special status in
this regard. The advantage of our method is that we get a
goodness-of-fit statistic in the likelihood which can be used to
compare models or judge whether a particular model is a good fit. Note
that the model is not just the cosmology, but includes modelling the
distributions of $x_1$ and $c$.

With this MLE, we can construct a confidence region in the
10-dimensional parameter space by defining its boundary as one of
constant $\mathcal{L}$. So long as we do not cross a boundary in
parameter space, this volume will asymptotically have the coverage
probability
\begin{equation}
  p_\textrm{cov} = \int_0^{-2\log{\mathcal{L}}/{\mathcal{L}}_\textrm{max}} 
  f_{\chi^2} (x; \nu) \textrm{d}x,
\label{eq:pcov}
\end{equation}
where $f_{\chi^2}(x;\nu)$ is the pdf of a chi-squared random variable
with $\nu$ degrees of freedom,
and $\mathcal{L}_\textrm{max}$ is the maximum likelihood. With 10
parameters in the present model, the values
$p_\textrm{cov} \simeq \{0.68\; (``1\sigma"), 0.95\; (``2\sigma")\}$
give
$-2\log{\mathcal L}/{\mathcal L}_\textrm{max} \simeq \{11.54, 18.61\}$
respectively.

To eliminate the so-called `nuisance parameters', we set similar
bounds on the profile likelihood. Writing the interesting parameters
as $\theta$ and nuisance parameters as $\phi$, the profile likelihood
is defined as
\begin{equation}
\mathcal{L}_\textrm{p} (\theta) = \max_\phi \mathcal{L} (\theta, \phi).
\end{equation}
We substitute $\mathcal{L}$ by $\mathcal{L}_\textrm{p}$ in
equation~\eqref{eq:pcov} in order to construct confidence regions in
this lower dimensional space; $\nu$ is now the dimension of the
remaining parameter space. Looking at the
$\Omega_\textrm{m}-\Omega_\Lambda$ plane, we have for
$p_\textrm{cov} \simeq \{0.68\; (``1\sigma"), 0.95\; (``2\sigma"),
0.997\; (``3 \sigma")\}$,
the values
$-2\log\mathcal L_\textrm{p}/\mathcal L_\textrm{max} \simeq \{2.30,
6.18, 11.8\}$ respectively.

\subsection{Comparison to other methods}

It is illuminating to relate our work to previously used methods in SN
Ia analyses. One method~\cite{Kim:2011hg} maximises a likelihood,
which is written in the case of uncorrelated magnitudes as
\begin{equation}
\tilde{\mathcal{L}} \propto \prod (2\pi \sigma_\textrm{tot}^2)^{-1/2}
 \exp\left(-\Delta\mu^2/2\sigma_\textrm{tot}^2 \right),
 \label{likenot}
\end{equation}
so it integrates over $\mu_\textrm{SN}$ to unity and can be used for
model comparison. From Equation~\eqref{LIKEINT} we see that this
corresponds to assuming \emph{flat} distributions for $x_1$ and
$c$. However the actual distributions of $\hat{x}_1$ and $\hat{c}$ are
close to gaussian, as seen in Fig.~\ref{fig:x1_c}. Moreover although
this likelihood apparently integrates to unity, it accounts for only
the $m_B^*$ data. Integration over the $x_1, c$ data demands compact
support for the flat distributions so the normalisation of the
likelihood becomes arbitrary, making model comparison tricky.

More commonly used~\cite{Perlmutter:1998np,Astier:2005qq} is the
`constrained $\chi^2$'
 
\begin{equation}
\chi^2 = \sum \Delta\mu^2/{(\sigma_\mu^2 + \sigma_\textrm{int}^2)},
\label{chi2}
\end{equation}
but this cannot be used to compare models, since it is \emph{tuned} to
be 1 per degree of freedom for the $\Lambda$CDM model by adjusting an
arbitrary error $\sigma_\textrm{int}$ added to each data point. This
has been criticised~\cite{Vishwakarma:2010nc,March:2011xa},
nevertheless the method continues to be widely used and the results
presented without emphasising that it is intended only for parameter
estimation for the \emph{assumed} ($\Lambda$CDM) model, rather than
determining if this is indeed the best model.

\begin{figure}[!h]
\begin{center}
    \includegraphics[width=\columnwidth]{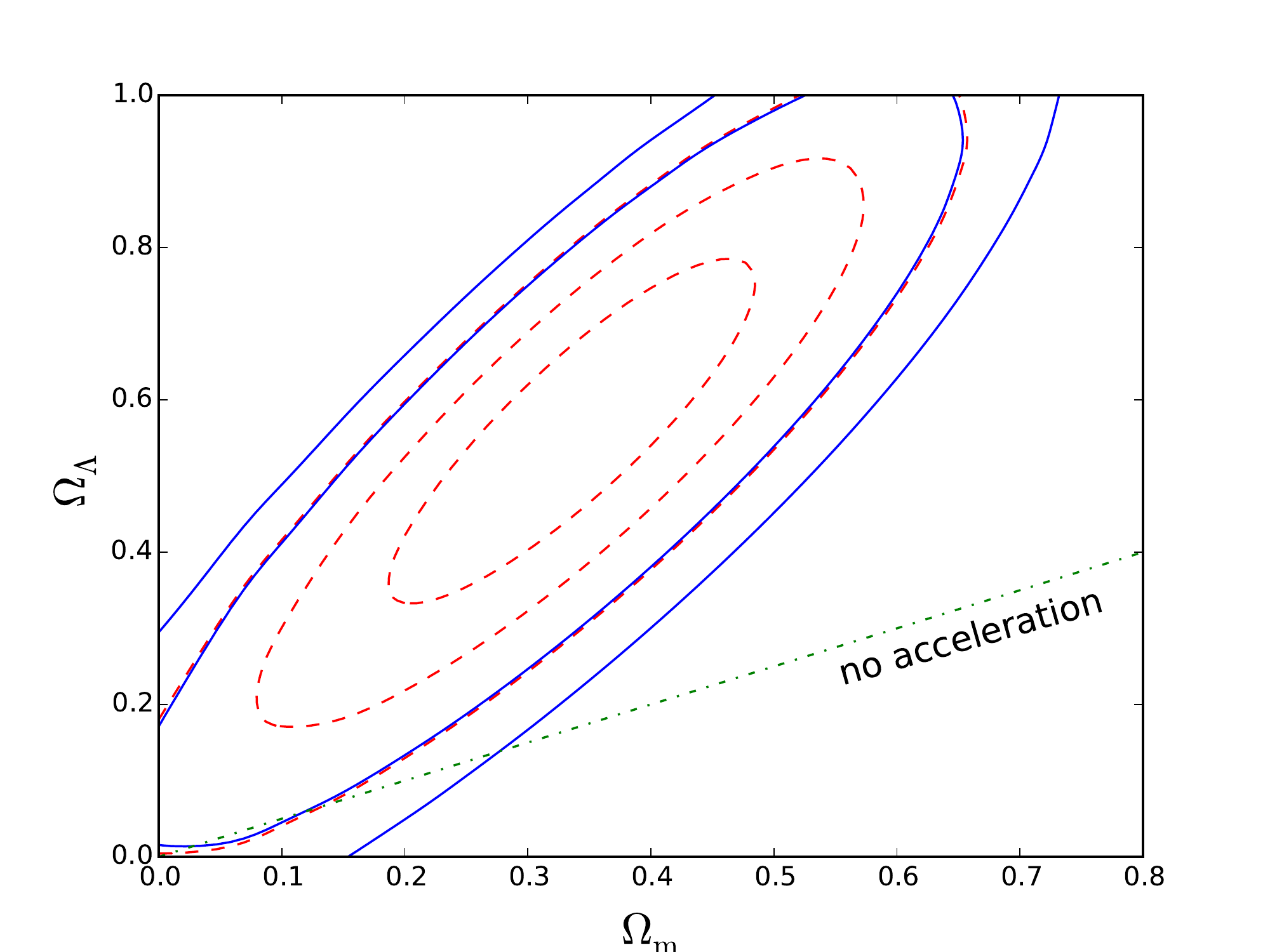}
    \caption{Contour plot of the profile likelihood in the
      $\Omega_\textrm{m}-\Omega_\Lambda$ plane. We show 1, 2 and
      3$\sigma$ contours, regarding all other parameters as nuisance
      parameters, as red dashed lines, while the blue lines are 1 and
      2$\sigma$ contours from the 10-dimensional parameter space
      projected on to this plane.}
\label{fig:contours}
\end{center}
\end{figure}

\begin{figure}[!t]
\begin{center}
    \includegraphics[width=\columnwidth]{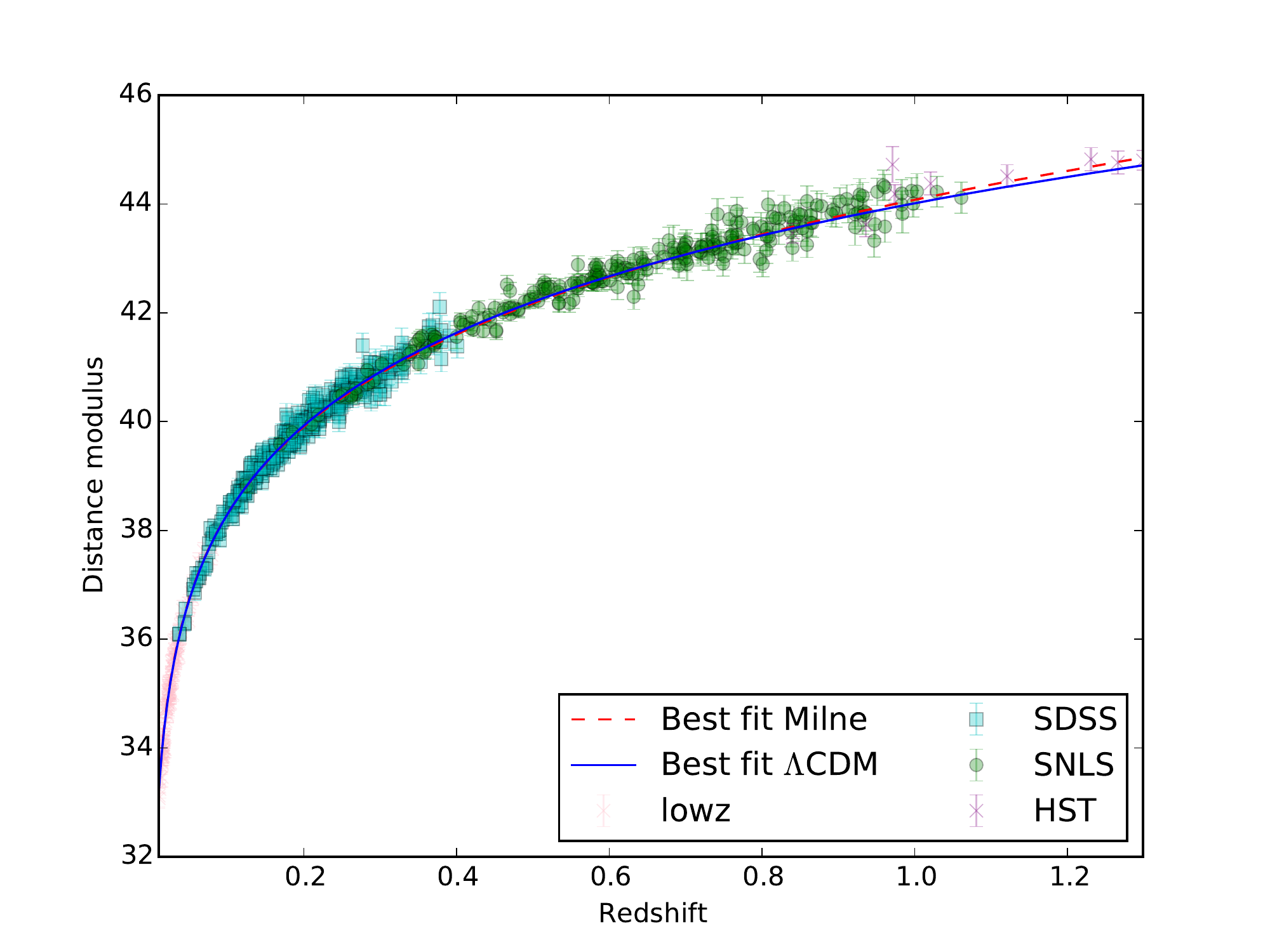}\\
    \includegraphics[width=\columnwidth]{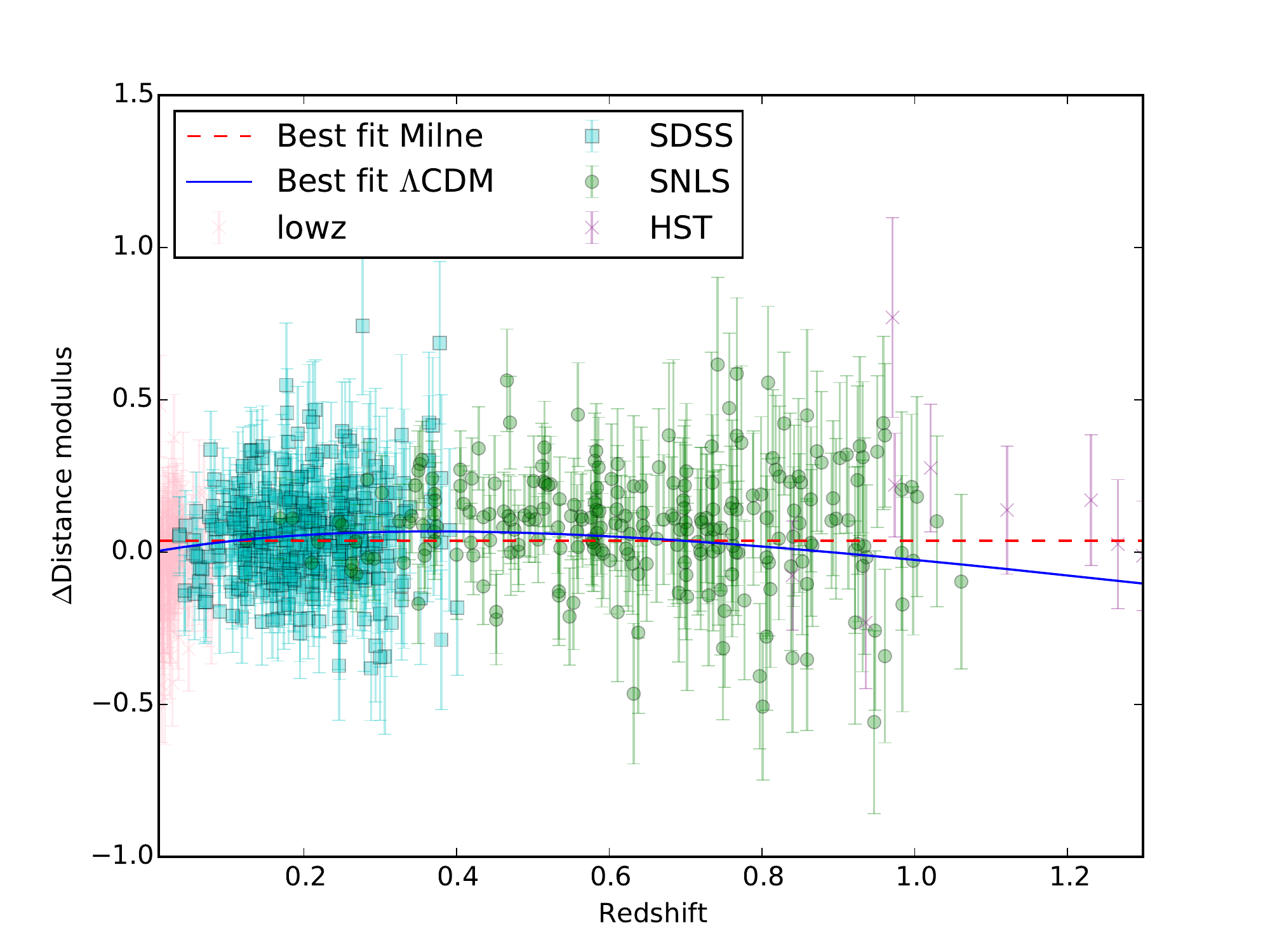}
    \caption{Comparison of the measured distance modulus with its
      expected value for the best fit accelerating universe
      ($\Lambda$CDM) and a universe expanding at constant velocity
      (Milne). The error bars include both experimental uncertainties
      and intrinsic dispersion. The bottom panel shows the residuals
      relative to the Milne model.}
\label{fig:hub}
\end{center}
\end{figure}

\section{Analysis of JLA catalogue}
\label{sec:an}

\begin{table*}[!h]
\centering
\begin{minipage}{160mm}
 \caption{Maximum likelihood parameters under specific
  constraints (in boldface).} ($-2\log\mathcal L_\textrm{max} = -214.97$)
  \label{tab:max}
 \begin{tabular}{@{}llcccccccccc}
Constraint & $-2\log\mathcal L/\mathcal{L}_\textrm{max}$ & $\Omega_\textrm{m}$ 
 & $\Omega_\Lambda$ & $\alpha$ & $x_{1,0}$ & $\sigma_{x_{1,0} }$ & $\beta$ 
 & $c_0$ & $\sigma_{c_0}$ & $M_0$ & $\sigma_{M_0}$ \\
	  \hline
None (best fit) & $\mathbf{0}$ & 0.341 & 0.569 & 0.134 & 0.038 & 0.932 
 & 3.059 & -0.016 & 0.071 & -19.052 & 0.108\\
Flat geometry & 0.147 & 0.376 & $\mathbf{0.624}$ & 0.135 & 0.039 & 0.932 
 & 3.060 &-0.016 & 0.071 & -19.055 & 0.108\\
Empty universe & 11.9 & $\mathbf{0.000}$ & $\mathbf{0.000}$ & 0.133 & 0.034 
 & 0.932 & 3.051 & -0.015 & 0.071 & -19.014 & 0.109 \\ 
Non-accelerating & 11.0 & 0.068 & $\mathbf{0.034}$ & 0.132 & 0.033 & 0.931 
 & 3.045 & -0.013 & 0.071 &-19.006 & 0.109 \\
Matter-less universe & 10.4 & $\mathbf{0.000}$ & 0.094 & 0.134 & 0.036 & 0.932 
 & 3.059 & -0.017 & 0.071 & -19.032 & 0.109 \\
Einstein-deSitter & 221.97 & $\mathbf{1.000}$ & $\mathbf{0.000}$ & 0.123 & 0.014 
 & 0.927 & 3.039 & 0.009 & 0.072 & -18.839 & 0.125
 \end{tabular}
\end{minipage}
\end{table*}

We focus on the Joint Lightcurve Analysis (JLA)
catalogue~\cite{Betoule:2014frx}. (All data used are available on
\protect\url{http://supernovae.in2p3.fr/sdss_snls_jla/ReadMe.html} ---
we use the \texttt{covmat\_v6}.) As shown already in
Fig.~\ref{fig:x1_c}, the distributions of the light curve fit
parameters $\hat{x}_1$ and $\hat{c}$ are well modelled as
gaussians. Maximisation of the likelihood under specific constraints
is summarised in Table~\ref{tab:max} and the profile likelihood
contours in the $\Omega_\textrm{m}-\Omega_\Lambda$ plane are shown in
Fig.~\ref{fig:contours}. In Fig.~\ref{fig:hub} we compare the measured
distance modulus,
$\hat{\mu_\textrm{SN}} = \hat{m}^*_B - M_0 + \alpha \hat {x}_1 -\beta
\hat{c}$
with its expected value in two cosmological models: `$\Lambda$CDM' is
the best fit accelerating universe while `Milne' is an universe
expanding with constant velocity. The error bars are the square root
of the diagonal elements of
$\Sigma_l + A^{\textrm{T}-1}\Sigma_\textrm{d} A^{-1}$ so include both
experimental uncertainties and intrinsic dispersion. We show also the
residuals with respect to the Milne model (which has been raised to
take into account the change in $M_0$).
      
\begin{figure}[h]
\centering
\includegraphics[width=\columnwidth]{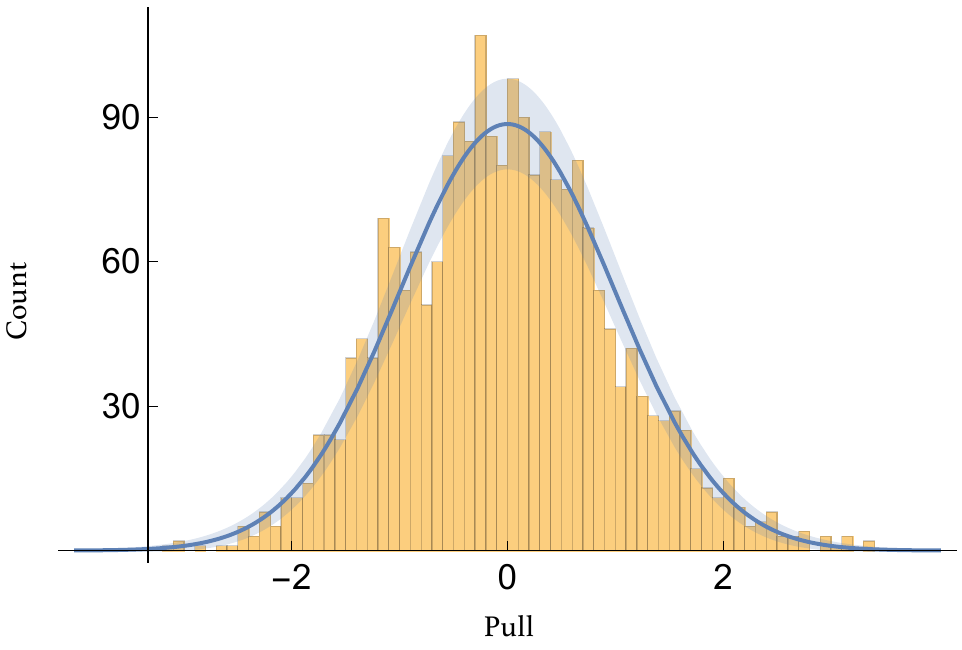}
\caption{Distribution of pulls \eqref{eq:pulls} for
  the best-fit model, compared to a normal distribution.}
  \label{fig:pulls}
\end{figure}

To assess how well our model describes the data, we present in
Fig.~\ref{fig:pulls} the `pull' distribution. These are defined as the
normalised, decorrelated residuals of the data,
\begin{equation}\label{eq:pulls}
\text{pulls} = (\hat{Z} - Y_0 A)U^{-1},
\end{equation}
where $U$ is the upper triangular Cholesky factor of the covariance
matrix $\Sigma_\textrm{d} + A^\textrm{T}\Sigma_l A$. Performing a K-S
test, comparing the pull distribution to a unit variance gaussian
gives a p-value of $0.1389$.

\begin{figure}[b]
\includegraphics[width=\columnwidth]{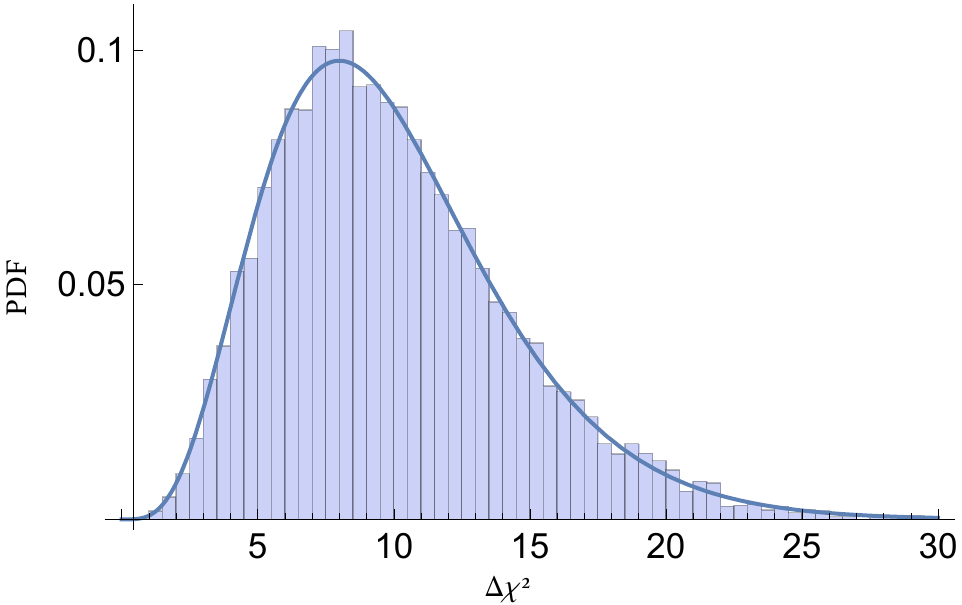}
\caption{The distribution of the likelihood ratio from Monte Carlo,
  with a $\chi^2$ distribution with 10 d.o.f. superimposed.}
\label{fig:MCchi2}
\end{figure}

To check the validity of our method and approximations, we do a Monte
Carlo simulation of experimental outcomes from a model with parameters
matching our best fit (see Table~\ref{tab:max}).
Figure~\ref{fig:MCchi2} shows the distribution of
$-2\log[\mathcal {L}_\textrm{true}/\mathcal{L}_\textrm{max}]$, which
is just as is expected.

\section{\label{sec:disc}Discussion}

That the SN Ia Hubble diagram appears consistent with an uniform rate
of expansion has been noted
earlier~\cite{Farley:2010gp,Melia:2012zy,Melia:2013hsa,Wei:2015xca}. We
have confirmed this by a rigorous statistical analysis, using the JLA
catalogue of 740 SN Ia processed by the SALT2 method. We find marginal
(i.e. $\lesssim 3\sigma$) evidence for the widely accepted claim that
the expansion of the universe is presently
accelerating~\cite{Goobar:2011iv}.

The Bayesian equivalent of this method (a ``Bayesian Hierarchical
Model") has been presented elsewhere~\cite{March:2011xa}.  We note
that a Bayesian consistency test~\cite{Marshall:2006zd} has been
applied (albeit using the flawed `likelihood' (equation \ref{likenot})
and `constrained $\chi^2$' (equation \ref{chi2}) methods) to determine
the consistency between the SN Ia data sets acquired with different
telescopes~\cite{Karpenka:2014moa}. These authors do find
inconsistencies in the UNION2 catalogue but none in JLA. This test had
been applied earlier to the UNION2.1 compilation finding no
contamination, but those authors~\cite{Heneka:2013hka} \emph{fixed}
the light curve fit `nuisance' parameters, so their result is
inconclusive. Including a `mass step' correction for the host galaxies
of SN~Ia~\cite{Betoule:2014frx} has little effect.

While our gaussian model \eqref{MODELGAUSS} is not perfect,
it appears to be an adequate first step towards understanding SN Ia
standardisation. One might be concerned that various selection effects
(e.g. Malmquist bias) affect the data. However to address this
adequately is beyond the scope of this work. We are concerned here
solely with performing the statistical analysis in an \emph{unbiased}
manner in order to highlight the different conclusion from previous
analyses~\cite{Betoule:2014frx} of the \emph{same} data.

We wish to emphasise that whether the expansion rate is accelerating
or not is a \emph{kinematic} test and it is simply for ease of
comparison with previous results that we choose to show the impact of
doing the correct statistical analysis in the usual $\Lambda$CDM
framework. In particular the `Milne model' should not be taken
literally to mean an empty universe since the deceleration due to
gravity can in principle be countered e.g. by bulk viscosity
associated with the formation of structure, resulting in expansion at
approximately constant velocity even in an universe containing matter
but no dark energy~\cite{Floerchinger:2014jsa}. Such a cosmology is
not \emph{prima facie} in conflict with observations of the angular
scale of fluctuations in the cosmic microwave background or of
baryonic acoustic oscillations, although this does require further
investigation. In any case, both of these are geometric rather than
dynamical measures and do not provide compelling \emph{direct}
evidence for a cosmological constant --- rather its value is inferred
from the assumed `cosmic sum rule':
$\Omega_\Lambda = 1 - \Omega_\textrm{m} + \Omega_k $. This would be
altered if additional terms due to the back reaction of
inhomogeneities are included in the Friedmann
equations~\cite{Buchert:2011sx}.

The CODEX experiment on the European Extremely Large Telescope aims to
measure the `redshift drift' over a 10-15 year period to determine
whether the expansion rate is really accelerating~\cite{Liske:2008ph}.

\newpage
\section*{Figure legends}

Fig.1: Distribution of the stretch and colour correction parameters in
the JLA sample~\cite{Betoule:2014frx}, with gaussians superimposed.

  Fig.2: Contour plot of the profile likelihood in the
  $\Omega_\textrm{m}-\Omega_\Lambda$ plane. We show 1, 2 and 3$\sigma$
  contours, regarding all other parameters as nuisance parameters, as
  red dashed lines, while the blue lines are 1 and 2$\sigma$ contours
  from the 10-dimensional parameter space projected on to this plane.

  Fig.3: Comparison of the measured distance modulus with its expected
  value for the best fit accelerating universe ($\Lambda$CDM) and a
  universe expanding at constant velocity (Milne). The error bars
  include both experimental uncertainties and intrinsic
  dispersion. The bottom panel shows the residuals relative to the
  Milne model.

Fig.4: Distribution of pulls \eqref{eq:pulls} for
  the best-fit model compared to a normal distribution.

Fig.5: The distribution of the likelihood ratio from Monte Carlo,
  with a $\chi^2$ distribution with 10 d.o.f. superimposed.

\section*{\label{APP:CONFIDENCE}Methods: Confidence ellipsoids}

The confidence ellipsoid is the collection of points
$x=\{\Omega_\text{m}, \Omega_\Lambda, \alpha, x_0, \sigma_{x_0}^2,
\beta, c_0, \sigma_{c_0}^2, M_0, \sigma_{M_0}^2 \}$, which obey
\begin{eqnarray}
[x - x_\text{MLE}] \mathcal{F} [x - x_\text{MLE}]^\text{T} \leq \Delta\chi^2,
\end{eqnarray}
where $\mathcal{F}$ is a symmetric matrix and $x_\text{MLE}$ is the
MLE. The enclosed volume is a confidence region with coverage
probability corresponding with high precision to the value obtained
from Equation~\eqref{eq:pcov}. The eigenvectors of $\mathcal F$ are
then the principal axes of the ellipsoid, and the eigenvalues are the
inverse squares of the lengths of the principal axes. We approximate
this matrix with the sample covariance from the MC of
section~\ref{sec:an} as $\mathcal{F} = \text{cov}(x, x)^{-1}$.

To make reading the matrix of eigenvectors easier, we round all
numbers to 0.1. Thus, we get the following approximate eigenvectors of
$\mathcal{F}$, in columns

\begin{equation}
\begin{array}{l}
\Omega_\text{m} \\ \Omega_\Lambda \\ \alpha \\ x_0 \\ \sigma_{x_0}^2 \\ 
 \beta \\ c_0 \\ \sigma_{c_0}^2 \\ M_0 \\ \sigma_{M_0}^2 \\
\end{array}
\begin{pmatrix}
0.5 & 0 & 0 & 0.8 & 0.1 & -0.2 & 0 & 0 & 0 & 0 \\
0.8 & 0 & 0 & -0.5 & -0.1 & 0.2 & 0 & 0 & 0 & 0 \\
0 & 0 & 0 & 0 & 0 & 0 & 1 & 0 & 0 & 0 \\
0 & 0 & 0 & -0.1 & 1 & 0 & 0 & 0 & 0 & 0 \\
0 & 0 & 1 & 0 & 0 & 0 & 0 & 0 & 0 & 0 \\
0 & 1 & 0 & 0 & 0 & 0 & 0 & 0 & 0 & 0 \\
0 & 0 & 0 & 0 & 0 & 0.1 & 0 & 1 & 0 & 0 \\
0 & 0 & 0 & 0 & 0 & 0 & 0 & 0 & 0 & 1 \\
-0.1 & 0 & 0 & 0.3 & 0.1 & 1 & 0 & 0.1 & 0 & 0 \\
0 & 0 & 0 & 0 & 0 & 0 & 0 & 0 & 1 & 0 \\
\end{pmatrix}
\end{equation}
with respective lengths of semi-axes

\begin{eqnarray}
10^{-3}\{172,& 85.1, 49.8, 43.9, 38.1, \nonumber\\ 
		9.89,& 5.93,4.24, 1.01, 0.304 \}
\end{eqnarray}
We also list the rounded correlation matrix,
\begin{equation}
\begin{pmatrix}
\Omega_\text{m} \\
 0.9 & \Omega_\Lambda  \\
 0 & 0 & \alpha \\
 0 & 0 & 0 & x_0 \\
 0 & 0 & -0.1 & 0 & \sigma_{x_0}^2  \\
 0 & 0 & 0 & 0 & 0 & \beta  \\
 0.1 & -0.1 & 0 & 0 & 0 & 0 & c_0 \\
 0 & 0 & 0 & 0 & 0 & -0.3 & 0 &  \sigma_{c_0}^2 \\
 -0.2 & -0.6 & 0 & 0 & 0 & 0.1 & 0.2 & 0 & M_0 \\
 0 & 0 & -0.1 & 0 & 0 & -0.3 & 0 & 0 & 0 & \sigma_{M_0}^2 \\
\end{pmatrix}
\end{equation}
We see that the only pronounced correlations are between
$\Omega_\text{m}, \Omega_\Lambda$ and $M_0$. This is also apparent
from Table~\ref{tab:max}.

\section*{\label{APP:CODE}Code availability}

The code and data used in the analysis are available at:
\protect\url{http://dx.doi.org/10.5281/zenodo.34487}

\section*{Acknowledgments}
We thank the JLA collaboration for making their data and software
public and M. Betoule for making the corrections we suggested to the
catalogue. This work was supported by the Danish National Research
Foundation through the Discovery Center at the Niels Bohr Institute
and the award of a Niels Bohr Professorship to S.S.

\section*{Contributions}
All authors participated in the analysis and in writing the paper.

\section*{Competing financial interests}
The authors have no competing financial interests.

\section*{Corresponding author}
Correspondence to: S. Sarkar


\begin{thebibliography}{99}
\expandafter\ifx\csname url\endcsname\relax
  \def\url#1{\texttt{#1}}\fi
\expandafter\ifx\csname urlprefix\endcsname\relax\def\urlprefix{URL }\fi
\providecommand{\bibinfo}[2]{#2}
\providecommand{\eprint}[2][]{\url{#2}}

\bibitem{Perlmutter:1998np}
\bibinfo{author}{Perlmutter, S.} \emph{et~al.}
\newblock \bibinfo{title}{{Measurements of Omega and Lambda from 42 high
  redshift supernovae}}.
\newblock \emph{\bibinfo{journal}{Astrophys.J.}}
  \textbf{\bibinfo{volume}{517}}, \bibinfo{pages}{565}
  (\bibinfo{year}{1999}).

\bibitem{Riess:1998cb}
\bibinfo{author}{Riess, A.~G.} \emph{et~al.}
\newblock \bibinfo{title}{{Observational evidence from supernovae for an
  accelerating universe and a cosmological constant}}.
\newblock \emph{\bibinfo{journal}{Astron.J.}} \textbf{\bibinfo{volume}{116}},
  \bibinfo{pages}{1009} (\bibinfo{year}{1998}).

\bibitem{Goobar:2011iv}
\bibinfo{author}{Goobar, A.} \& \bibinfo{author}{Leibundgut, B.}
\newblock \bibinfo{title}{{Supernova cosmology: legacy and future}}.
\newblock \emph{\bibinfo{journal}{Ann.Rev.Nucl.Part.Sci.}}
  \textbf{\bibinfo{volume}{61}}, \bibinfo{pages}{251}
  (\bibinfo{year}{2011}).

\bibitem{Phillips:1993ng}
\bibinfo{author}{Phillips, M.}
\newblock \bibinfo{title}{{The absolute magnitudes of type Ia supernovae}}.
\newblock \emph{\bibinfo{journal}{Astrophys.J.}}
  \textbf{\bibinfo{volume}{413}}, \bibinfo{pages}{L105}
  (\bibinfo{year}{1993}).

\bibitem{Tripp:1997wt}
\bibinfo{author}{Tripp, R.}
\newblock \bibinfo{title}{{A two-parameter luminosity correction for type Ia
  supernovae}}.
\newblock \emph{\bibinfo{journal}{Astron.Astrophys.}}
  \textbf{\bibinfo{volume}{331}}, \bibinfo{pages}{815}
  (\bibinfo{year}{1998}).

\bibitem{Kelly:2009iy}
\bibinfo{author}{Kelly, P.~L.}, \bibinfo{author}{Hicken, M.},
  \bibinfo{author}{Burke, D.~L.}, \bibinfo{author}{Mandel, K.~S.} \&
  \bibinfo{author}{Kirshner, R.~P.}
\newblock \bibinfo{title}{{Hubble residuals of nearby type Ia supernovae are
  correlated with host galaxy masses}}.
\newblock \emph{\bibinfo{journal}{Astrophys.J.}}
  \textbf{\bibinfo{volume}{715}}, \bibinfo{pages}{743}
  (\bibinfo{year}{2010}).

\bibitem{Hayden:2012aa}
\bibinfo{author}{Hayden, B.~T.} \emph{et~al.}
\newblock \bibinfo{title}{{The fundamental metallicity relation reduces type Ia
  SN Hubble residuals more than host mass alone}}.
\newblock \emph{\bibinfo{journal}{Astrophys.J.}}
  \textbf{\bibinfo{volume}{764}}, \bibinfo{pages}{191} (\bibinfo{year}{2013}).

\bibitem{Astier:2005qq}
\bibinfo{author}{Astier, P.} \emph{et~al.}
\newblock \bibinfo{title}{{The supernova legacy survey: Measurement of
  $\Omega_M$, $\Omega_\Lambda$ and $w$ from the first year data set}}.
\newblock \emph{\bibinfo{journal}{Astron.Astrophys.}}
  \textbf{\bibinfo{volume}{447}}, \bibinfo{pages}{31}
  (\bibinfo{year}{2006}).

\bibitem{Conley:2006qb}
\bibinfo{author}{Conley, A.~J.} \emph{et~al.}
\newblock \bibinfo{title}{{Measurement of Omega(m), Omega(lambda) from a blind
  analysis of type Ia supernovae with CMAGIC: using colour information to verify
  the acceleration of the Universe}}.
\newblock \emph{\bibinfo{journal}{Astrophys.J.}}
  \textbf{\bibinfo{volume}{644}}, \bibinfo{pages}{1}
  (\bibinfo{year}{2006}).

\bibitem{Kowalski:2008ez}
\bibinfo{author}{Kowalski, M.} \emph{et~al.}
\newblock \bibinfo{title}{{Improved cosmological constraints from new, old and
  combined supernova datasets}}.
\newblock \emph{\bibinfo{journal}{Astrophys.J.}}
  \textbf{\bibinfo{volume}{686}}, \bibinfo{pages}{749}
  (\bibinfo{year}{2008}).

\bibitem{Betoule:2014frx}
\bibinfo{author}{Betoule, M.} \emph{et~al.}
\newblock \bibinfo{title}{{Improved cosmological constraints from a joint
  analysis of the SDSS-II and SNLS supernova samples}}.
\newblock \emph{\bibinfo{journal}{Astron.Astrophys.}}
  \textbf{\bibinfo{volume}{568}}, \bibinfo{pages}{A22} (\bibinfo{year}{2014}).

\bibitem{Vishwakarma:2010nc}
\bibinfo{author}{Vishwakarma, R.~G.} \& \bibinfo{author}{Narlikar, J.~V.}
\newblock \bibinfo{title}{{A critique of supernova data analysis in
  cosmology}}.
\newblock \emph{\bibinfo{journal}{Res.Astron.Astrophys.}}
  \textbf{\bibinfo{volume}{10}}, \bibinfo{pages}{1195}
  (\bibinfo{year}{2010}).

\bibitem{March:2011xa}
\bibinfo{author}{March, M.}, \bibinfo{author}{Trotta, R.},
  \bibinfo{author}{Berkes, P.}, \bibinfo{author}{Starkman, G.} \&
  \bibinfo{author}{Vaudrevange, P.}
\newblock \bibinfo{title}{{Improved constraints on cosmological parameters from
  SN Ia data}}.
\newblock \emph{\bibinfo{journal}{Mon.Not.Roy.Astron.Soc.}}
  \textbf{\bibinfo{volume}{418}}, \bibinfo{pages}{2308}
  (\bibinfo{year}{2011}).

\bibitem{Kim:2011hg}
\bibinfo{author}{Kim, A.}
\newblock \bibinfo{title}{{Type Ia supernova intrinsic magnitude dispersion and
  the fitting of cosmological parameters}}.
\newblock \emph{\bibinfo{journal}{Publ.Astron.Soc.Pac.}}
  \textbf{\bibinfo{volume}{123}}, \bibinfo{pages}{230} (\bibinfo{year}{2011}).

\bibitem{Lago:2011pk}
\bibinfo{author}{Lago, B.} \emph{et~al.}
\newblock \bibinfo{title}{{Type Ia supernova parameter estimation: a comparison
  of two approaches using current datasets}}.
\newblock \emph{\bibinfo{journal}{Astron.Astrophys.}}
  \textbf{\bibinfo{volume}{541}}, \bibinfo{pages}{A110} (\bibinfo{year}{2012}).

\bibitem{Wei:2015xca}
\bibinfo{author}{Wei, J.-J.}, \bibinfo{author}{Wu, X.-F.},
  \bibinfo{author}{Melia, F.} \& \bibinfo{author}{Maier, R.~S.}
\newblock \bibinfo{title}{{A comparative analysis of the supernova legacy
  survey sample with $\Lambda$CDM and the $R_h = ct$ Universe}}.
\newblock \emph{\bibinfo{journal}{Astron.J.}} \textbf{\bibinfo{volume}{149}},
  \bibinfo{pages}{102} (\bibinfo{year}{2015}).

\bibitem{Hicken:2009dk}
\bibinfo{author}{Hicken, M.} \emph{et~al.}
\newblock \bibinfo{title}{{Improved dark energy constraints from $\sim$100 new CfA
 supernova Type Ia light curves}}.
\newblock \emph{\bibinfo{journal}{Astrophys.J.}}
  \textbf{\bibinfo{volume}{700}}, \bibinfo{pages}{1097}
  (\bibinfo{year}{2009}).

\bibitem{Guy:2005me}
\bibinfo{author}{Guy, J.}, \bibinfo{author}{Astier, P.},
  \bibinfo{author}{Nobili, S.}, \bibinfo{author}{Regnault, N.} \&
  \bibinfo{author}{Pain, R.}
\newblock \bibinfo{title}{{SALT: A spectral adaptive light curve template for
  Type Ia supernovae}}.
\newblock \emph{\bibinfo{journal}{Astron.Astrophys.}}
  \textbf{\bibinfo{volume}{443}}, \bibinfo{pages}{781}
  (\bibinfo{year}{2005}).

\bibitem{Guy:2007dv}
\bibinfo{author}{Guy, J.} \emph{et~al.}
\newblock \bibinfo{title}{{SALT2: Using distant supernovae to improve the use
  of Type Ia supernovae as distance indicators}}.
\newblock \emph{\bibinfo{journal}{Astron.Astrophys.}}
  \textbf{\bibinfo{volume}{466}}, \bibinfo{pages}{11}
  (\bibinfo{year}{2007}).

\bibitem{Hoflich:1996sx}
\bibinfo{author}{Hoflich, P.} \emph{et~al.}
\newblock \bibinfo{title}{{Maximum brightness and post-maximum decline of light
  curves of SN Ia: a comparison of theory and observations}}.
\newblock \emph{\bibinfo{journal}{Astrophys.J.}}
  \textbf{\bibinfo{volume}{472}}, \bibinfo{pages}{L81} (\bibinfo{year}{1996}).

\bibitem{Kasen:2006is}
\bibinfo{author}{Kasen, D.} \& \bibinfo{author}{Woosley, S.}
\newblock \bibinfo{title}{{On the origin of the type Ia supernova
  width-luminosity relation}}.
\newblock \emph{\bibinfo{journal}{Astrophys.J.}}
  \textbf{\bibinfo{volume}{656}}, \bibinfo{pages}{661}
  (\bibinfo{year}{2007}).

\bibitem{Farley:2010gp}
\bibinfo{author}{Farley, F.~J.}
\newblock \bibinfo{title}{{Does gravity operate between galaxies? Observational
  evidence re-examined}}.
\newblock \emph{\bibinfo{journal}{Proc.Roy.Soc.Lond.}}
  \textbf{\bibinfo{volume}{A466}}, \bibinfo{pages}{3089}
  (\bibinfo{year}{2010}).

\bibitem{Melia:2012zy}
\bibinfo{author}{Melia, F.}
\newblock \bibinfo{title}{{Fitting the Union2.1 SN sample with the $R_h=ct$
  Universe}}.
\newblock \emph{\bibinfo{journal}{Astron.J.}} \textbf{\bibinfo{volume}{144}},
  \bibinfo{pages}{110} (\bibinfo{year}{2012}).

\bibitem{Melia:2013hsa}
\bibinfo{author}{Melia, F.} \& \bibinfo{author}{Maier, R.~S.}
\newblock \bibinfo{title}{{Cosmic chronometers in the $R_h=ct$ Universe}}.
\newblock \emph{\bibinfo{journal}{Mon.Not.Roy.Astron.Soc.}}
  \textbf{\bibinfo{volume}{432}}, \bibinfo{pages}{2669}
  (\bibinfo{year}{2013}).

\bibitem{Marshall:2006zd}
\bibinfo{author}{Marshall, P.}, \bibinfo{author}{Rajguru, N.} \&
  \bibinfo{author}{Slosar, A.}
\newblock \bibinfo{title}{{Bayesian evidence as a tool for comparing
  datasets}}.
\newblock \emph{\bibinfo{journal}{Phys.Rev.}} \textbf{\bibinfo{volume}{D73}},
  \bibinfo{pages}{067302} (\bibinfo{year}{2006}).

\bibitem{Karpenka:2014moa}
\bibinfo{author}{Karpenka, N.}, \bibinfo{author}{Feroz, F.} \&
  \bibinfo{author}{Hobson, M.}
\newblock \bibinfo{title}{{Testing the mutual consistency of different
  supernovae surveys}}.
\newblock \emph{\bibinfo{journal}{Mon.Not.Roy.Astron.Soc.}}
  \textbf{\bibinfo{volume}{449}}, \bibinfo{pages}{2405} (\bibinfo{year}{2015}).

\bibitem{Heneka:2013hka}
\bibinfo{author}{Heneka, C.}, \bibinfo{author}{Marra, V.} \&
  \bibinfo{author}{Amendola, L.}
\newblock \bibinfo{title}{{Extensive search for systematic bias in supernova Ia
  data}}.
\newblock \emph{\bibinfo{journal}{Mon.Not.Roy.Astron.Soc.}}
  \textbf{\bibinfo{volume}{439}}, \bibinfo{pages}{1855}
  (\bibinfo{year}{2014}).

\bibitem{Floerchinger:2014jsa}
\bibinfo{author}{Floerchinger, S.}, \bibinfo{author}{Tetradis, N.} \&
  \bibinfo{author}{Wiedemann, U.~A.}
\newblock \bibinfo{title}{{Accelerating cosmological expansion from shear and
  bulk viscosity}}.
\newblock \emph{\bibinfo{journal}{Phys. Rev. Lett.}}
  \textbf{\bibinfo{volume}{114}}, \bibinfo{pages}{091301}
  (\bibinfo{year}{2015}).

\bibitem{Buchert:2011sx}
\bibinfo{author}{Buchert, T.}, \bibinfo{author}{R\"as\"anen, S.} 
\newblock \bibinfo{title}{{Backreaction in late-time cosmology}}.
\newblock \emph{\bibinfo{journal}{Ann. Rev. Nucl. Part. Sci.}}
  \textbf{\bibinfo{volume}{62}}, \bibinfo{pages}{57}
  (\bibinfo{year}{2012}).


\bibitem{Liske:2008ph}
\bibinfo{author}{Liske, J.} \emph{et~al.}
\newblock \bibinfo{title}{{Cosmic dynamics in the era of extremely large
  telescopes}}.
\newblock \emph{\bibinfo{journal}{Mon. Not. Roy. Astron. Soc.}}
  \textbf{\bibinfo{volume}{386}}, \bibinfo{pages}{1192}
  (\bibinfo{year}{2008}).

\end{thebibliography}
\end{document}